\documentclass{webofc}
\usepackage[varg]{txfonts}   % Web of Conferences font
\begin{document}
\title{Towards the determination of the 3-dimensional structure of the proton using lattice QCD simulations}

\author{\firstname{Constantia} \lastname{Alexandrou}\inst{1,2}\fnsep\thanks{ Speaker, \email{alexand@ucy.ac.cy}}
  \and
      \firstname{Simone} \lastname{Bacchio}\inst{2}\fnsep\thanks{\email{s.bacchio@cyi.ac.cy}}
%             author if necessary}} \and
 %       \firstname{Third author} \lastname{Third author}\inst{3}\fnsep\thanks{\e%mail{Mail address for last
%             author if necessary}}
        % etc.
}

\institute{Department of Physics, University of Cyprus, PO Box 20537, 1678 Nicosia, Cyprus
\and
Computation-based Science and Technology Research Center, The Cyprus Institute, 20 Konstantinou Kavafi str., 2121 Aglantzia, Cyprus
%\and
%           Last address
          }

\abstract{%
 State-of-the-art lattice QCD simulations enable  the evaluation of nucleon  form factors and Mellin moments with controlled systematics, yielding results with   unprecedented accuracy. At the same time, new  theoretical approaches are allowing the direct computation of nucleon generalized parton distributions. We review recent lattice QCD results on these quantities that are paving the way for extracting a wealth of information on the 3-dimensional structure of the nucleon. 
}\vspace*{-0.3cm}
\maketitle
\vspace*{-0.6cm}

\section{Introduction}\vspace*{-0.2cm}
\label{intro}
The Langragian of the theory of the strong interaction, Quantum Chromodynamics (QCD), was written 50 years ago~\cite{Fritzsch:1973pi}. Solving it and obtaining quantitative results is very challenging due to the non-perturbative nature of the strong interactions. The lattice formulation provides the non-perturbative framework to regularised the theory~\cite{Wilson:1974sk} and, after rotating to imaginary time,  to compute hadronic properties using {\it ab initio} simulations~\cite{Luscher:1998pe,Luscher:2002pz}.
The development of powerful computers have helped immensely the field. However, equally important were theoretical and algorithmic breakthroughs~\cite{Luscher:2003vf,Luscher:2003qa,Luscher:2004pav}. Nowadays, simulations are carried out at physical values of the quark masses, large enough lattice size and small enough lattice spacing. New theoretical developments are extending the quantities that can be extracted from such simulations, opening new areas where lattice QCD can provide insights and valuable  input to phenomenology and  experiments. A particularly relevant theoretical development for determining the structure of the nucleon is the large momentum effective theory (LaMET)~\cite{Ji:2013dva} that gives access to the  direct computation of parton distribution functions~(PDFs)  and their generalization, the generalized parton distributions (GPDs). \vspace*{-0.3cm}
\section{State-of-the lattice QCD simulations} 
\begin{figure}[h!]
    \centering
    \includegraphics[width=0.6\linewidth]{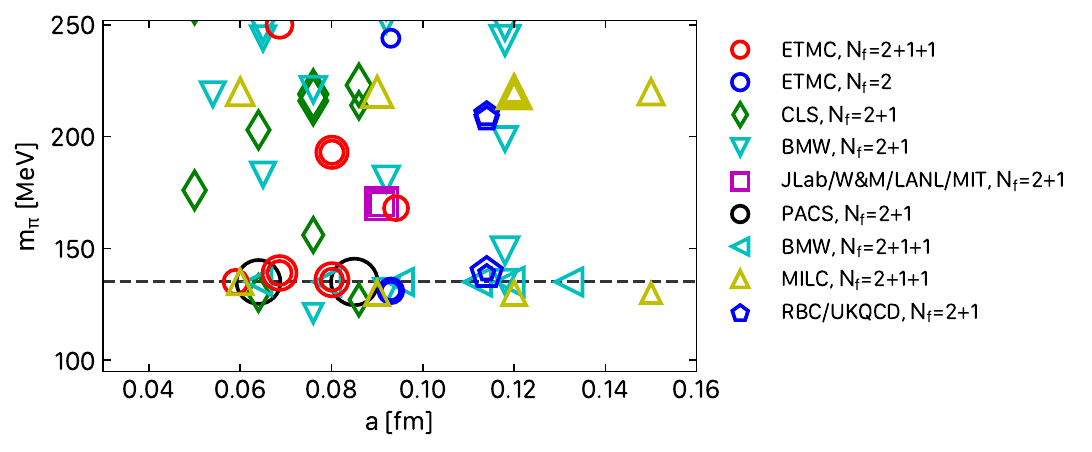}
    \includegraphics[width=0.33\linewidth]{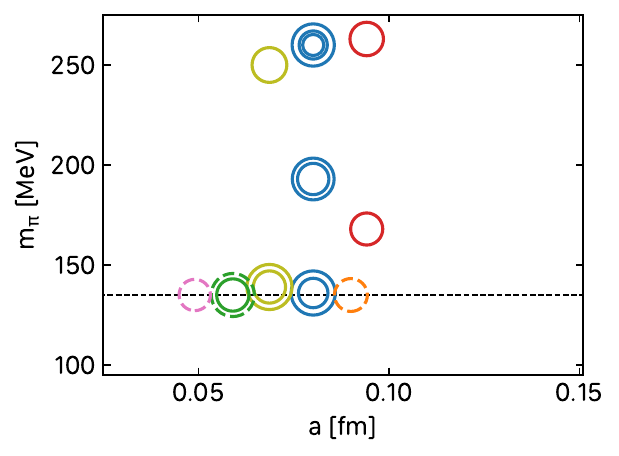}\vspace*{-0.3cm}
    \caption{Left: Gauge ensembles by various collaborations using clover improved fermions (ETMC, CLS, BMW, JLab/W\&M/LANL/MIT, PACS), staggered fermions (BMW, MILC) and Domain wall fermions (RBC/UKQCD). The size of the symbols reflects the spatial volume of the lattice. Right: Gauge ensembles produced (sold circles) or under production (dashed circles) by ETMC.  The dashed horizintal lines show the physical value of the pion mass.}
    \label{fig:simulations}\vspace*{-0.5cm}
\end{figure}

\vspace*{-0.2cm}
In Fig.~\ref{fig:simulations} we show a summary of the zero-temperature  gauge ensembles available for hadron structure studies. The three most used discretization fermion schemes are clover-type, staggered and domain wall  fermions. All major collaborations have gauge ensembles simulated approximately at the  physical value of the pion mass. The Extended Twisted Mass Collaboration (ETMC) completed the simulation of three gauge ensembles at three different lattice spacings  using twisted mass clover-improved fermions with the masses of the  light, strange and charm quarks tuned close to their physical values. These three ensembles enable the extraction of form factors and Mellin moments  taking the continuum limit directly at the physical pion mass, eliminating the need for chiral extrapolations.

\begin{table}[t!]
    \centering
    \begin{tabular}{c|c|c|c|c|c}
    \hline\hline
       Ensemble  & $V/a^4$ & $\beta$ & $a$ [fm] & $m_\pi$ [MeV]  & $m_\pi L$ \\
       \hline 
        \texttt{cB211.072.64} & $64^3 \times 128$ & 1.778 &  0.07957(13) & 140.2(2) & 3.62 \\
        \texttt{cC211.060.80} & $80^3 \times 160$ & 1.836 &  0.06821(13) & 136.7(2) & 3.78 \\
        \texttt{cD211.054.96} & $96^3 \times 192$ & 1.900 &  0.05692(12) & 140.8(2) & 3.90 \\
        \hline
    \end{tabular}
    \caption{Parameters for the $N_f=2+1+1$ ETMC  ensembles. In the first column, we give the name of the ensemble, in the second the lattice volume, in the third $\beta=6/g^2$ with $g$  the bare coupling constant, in the fourth the lattice spacing, in the fifth the pion mass, and in the sixth the value of $m_\pi L$. Lattice spacings and pion masses are taken from Ref.~\cite{ExtendedTwistedMass:2022jpw}.} 
    \label{tab:ens}\vspace*{-0.8cm}
\end{table}

\section{Axial form factors} 
\begin{figure}[h!]
    \begin{minipage}{0.5\linewidth}
      \includegraphics[width=\linewidth]{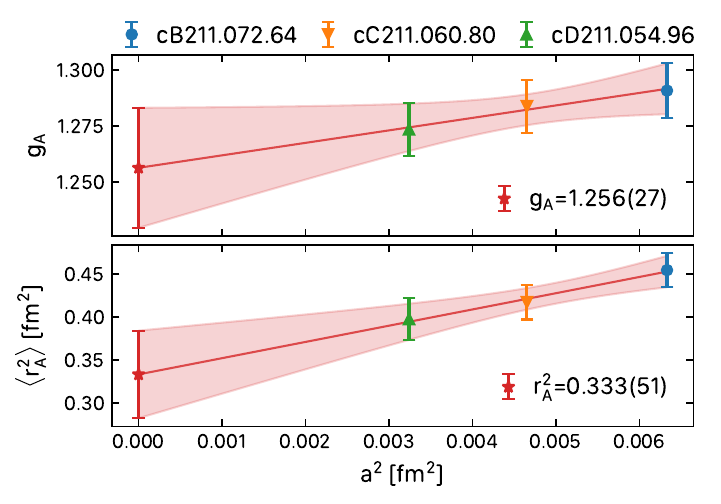}
    \end{minipage}\hfill 
     \begin{minipage}{0.4\linewidth}\vspace*{-0.5cm}
       \includegraphics[width=\linewidth]{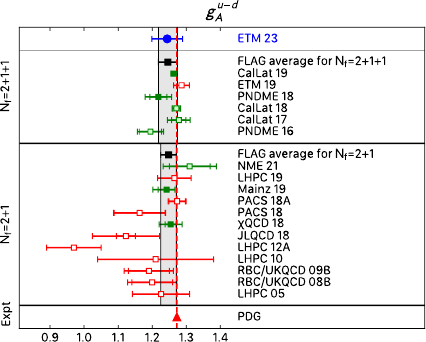}
       \end{minipage}\vspace*{-0.3cm}
    \caption{Left: Continuum extrapolation of $g_A$ and the axial radius
      $r_A$ using the the three ETMC ensembles of Table~\ref{tab:ens}~\cite{Alexandrou:2023qbg}. Right: Comparison of lattice QCD results on $g_A$ as published by FLAG~\cite{FlavourLatticeAveragingGroupFLAG:2021npn} where we include  the ETMC result with the blue circle, labeled ETM 23. }
    \label{fig:gA}\vspace*{-0.6cm}
\end{figure}

\vspace*{-0.2cm}
 The nucleon axial form factors are important quantities for
weak interactions, neutrino scattering,  and parity violation experiments.
 Since the value of the nucleon axial charge $g_A$  is well known, it has been used over the years as a benchmark quantity for lattice QCD computations. The nucleon axial charge, $g_A$, is extracted directly in lattice QCD from the forward matrix element of the axial-vector current. In Fig.~\ref{fig:gA}, we show the continuum  limit using the three ETMC ensembles of Table~\ref{tab:ens}. In the continuum limit, we find $ g_A=1.245(28)(14)$, where the first error is statistical and the second the systematic due to excited states.  In Fig.~\ref{fig:gA}, we also show results of numerous lattice QCD studies as presented in the 2021 FLAG report~\cite{FlavourLatticeAveragingGroupFLAG:2021npn}. As can be seen, the lattice results reproduce the value of $g_A$. 

\begin{figure}[h!]
    \centering
    \includegraphics[width=0.46\linewidth]{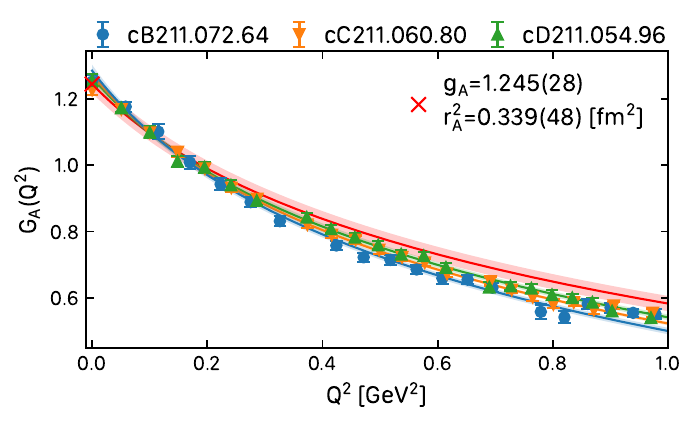}
    \includegraphics[width=0.46\linewidth]{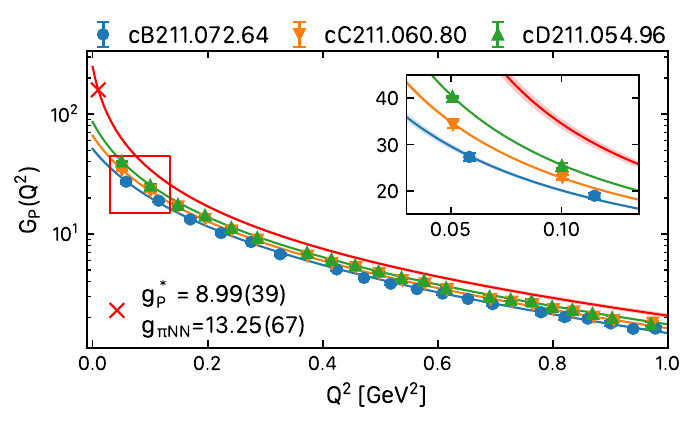} \vspace*{-0.3cm}
    \caption{ Results on the axial  $G_A(Q^2)$ (left) and induced pseudoscalar $G_P(Q^2)$ (right) form factors as a function of $Q^2$. The red solid line shows the continuum extrapolation~\cite{Alexandrou:2023qbg}. }
    \label{fig:FFs}\vspace*{-0.6cm}
\end{figure}

\begin{figure}[h!]
    \centering
    \includegraphics[width=0.46\linewidth]{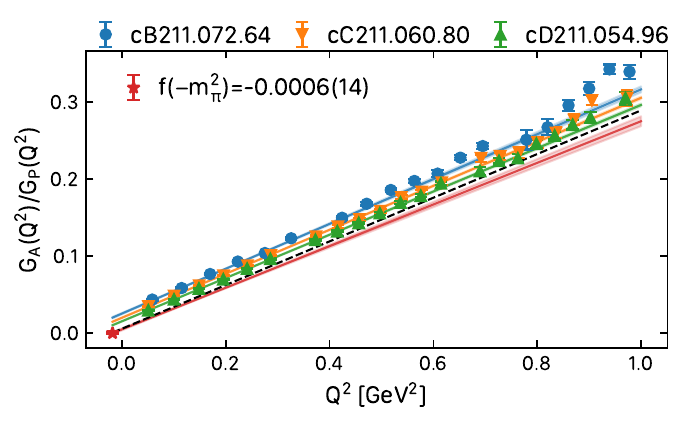}
    \includegraphics[width=0.48\linewidth]{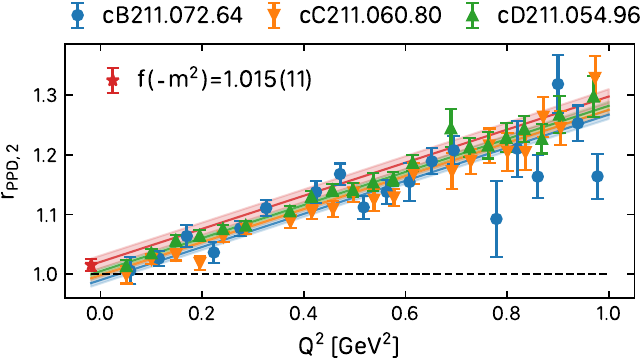}
    \caption{ Results on the ratios  $G_A(Q^2)/G_P(Q^2)$ (left) and $r_{PPD,2}$  (right)  as a function of $Q^2$.  The red bands
show the continuum extrapolation~\cite{Alexandrou:2023qbg}. }
    \label{fig:ratios}\vspace*{-0.3cm}
\end{figure}

Results for the axial form factors are shown in Fig.~\ref{fig:FFs}. We fit the $Q^2$-dependence using a dipole form or the z-expansion. From the slope in the continuum limit, we find for the axial radius  $\langle r^2_A \rangle=0.339(67)(06)~{\rm fm}^2$. We also check the pion pole dominance (PPD) hypothesis that predicts that the value of the  ratio
$ G_A(Q^2)/G_P(Q^2)|_{Q^2\rightarrow -m_\pi^2}=\frac{Q^2+m_\pi^2}{4m_N^2}$, close to the pole. We take the continuum limit using the Ansatz $f(Q^2,a^2)=c_0+c_1Q^2+c_2a^2+c_3a^2Q^2$. We find that, in the continuum limit, the slope $c_1\sim 1/4m_N^2$ and $c_0$ is consistent with zero, showing that PPD hypothesis is fulfilled, as shown in Fig.~\ref{fig:ratios}.  A quantity of interest for the induced pseudoscalar form factor is the induced pseudoscalar coupling determined at the muon capture point~\cite{Egger:2016hcg},  namely
$
     g_P^* \equiv \frac{m_\mu}{2 m_N} G_P(0.88\,m_\mu^2)
$
with $m_\mu=105.6$~MeV the muon mass. We find  that $g_P^*=8.99(39)(49)$, which agrees with   the value  of $8.44(16)$ obtained in chiral perturbation theory ~\cite{Bernard:1994wn}.
For a non-zero pion mass, the spontaneous breaking of chiral symmetry relates the axial-vector current to the pion field $\psi_\pi$, through the relation
$
    \partial^\mu A_\mu = F_\pi m_\pi^2 \psi_\pi.
$
In QCD, the axial Ward-Takahashi identity leads to the  partial conservation of the axial-vector current (PCAC) 
$
  \partial^\mu A_\mu= 2 m_q P,
 $
where $P$ is the pseudoscalar operator and $m_q=m_u=m_d$ is the light quark mass for degenerate up and down quarks, where $P=\bar{u} \gamma_5  u - \bar{d}  \gamma_5 d$ is the isovector pseudoscalar current.
The PCAC relation at the form factors level relates the axial and induced pseudoscalar form factors to the pseudoscalar form factor via the relation
\begin{equation}
    G_A(Q^2) - \frac{Q^2}{4 m_N^2} G_P(Q^2) = \frac{m_q}{m_N} G_5(Q^2).
    \label{Eq:PCAC_FFs}
\end{equation}
Using the PCAC relation,  it also follows that  the pion field can be expressed as
   $ \psi_\pi = \frac{2 m_q P}{F_\pi m_\pi^2}$
and  one can connect the pseudoscalar form factor to the pion-nucleon form factor $G_{\pi NN}(Q^2)$ as follows\cite{Alexandrou:2023qbg}
\begin{equation}
    m_qG_5(Q^2) = \frac{F_\pi m_\pi^2}{m_\pi^2 + Q^2} G_{\pi NN}(Q^2),
    \label{Eq:PtoPiNN}
\end{equation}
which is written so that it illustrates the pole structure of $G_5(Q^2)$ and the preferred usage of $m_qG_5(Q^2)$, which is a scale-independent quantity unlike $G_5(Q^2)$.  Substituting $m_qG_5(Q^2)$ in Eq.~(\ref{Eq:PCAC_FFs}),  one obtains the Goldberger-Treiman relation~\cite{Alexandrou:2007hr,Alexandrou:2007eyf}
\begin{equation}
G_A(Q^2)-\frac{Q^2}{4m_N^2} G_P(Q^2)=\frac{F_\pi m_\pi^2}{m_N(m^2_\pi+Q^2)}G_{\pi NN}(Q^2).
  \label{GT}    
\end{equation}

 \begin{figure}[h!]
   \begin{minipage}{0.33\linewidth}
     \includegraphics[width=\linewidth]{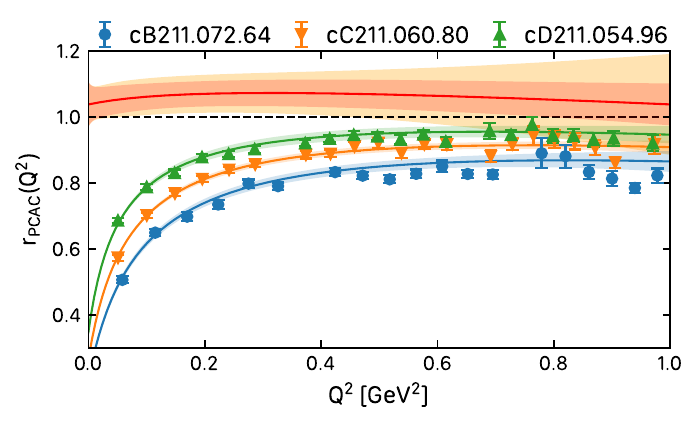}
   \end{minipage}\hfill
    \begin{minipage}{0.22\linewidth}
      \includegraphics[width=\linewidth]{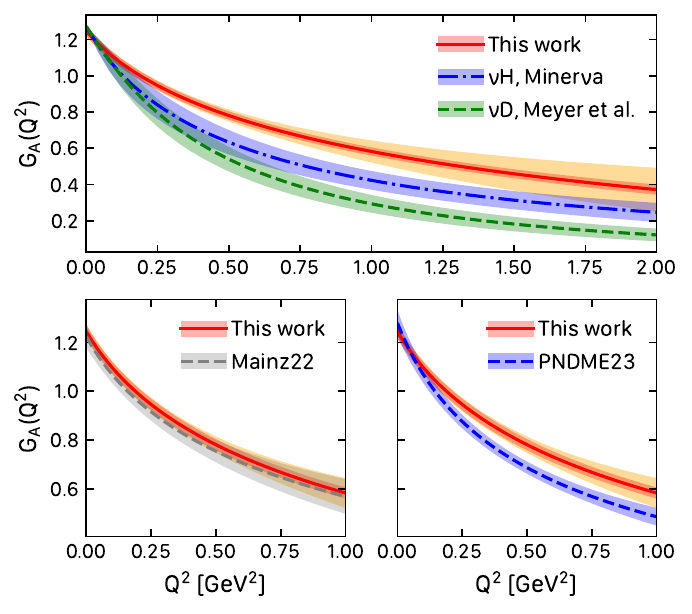}
      \end{minipage}\hfill
    \begin{minipage}{0.40\linewidth}
      \includegraphics[width=\linewidth]{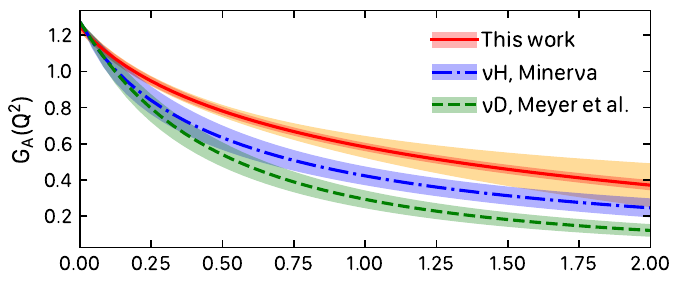}
      \end{minipage} \vspace*{-0.3cm}
     \caption{Left: $r_{\rm PCAC}$  for the three ETMC ensembles of Table~\ref{tab:ens} and in the continuum limit (red band) and when including the systematic error due to excited states (yellow band). The black dashed line is the expected result if PPD is satisfied at any $Q^2$.  Middle: Comparison of ETMC results with a recent computation by the CLS-Mainz group~\cite{Djukanovic:2022wru} shown with the gray dashed line with its error band. %and ii)  by PNDME~\cite{Jang:2023zts} shown with the blue dashed line with its error band.
       Right: Comparison of ETMC continuum values  on $G_A(Q^2)$ with  the fit to the deuterium bubble-chamber data~\cite{Meyer:2016oeg} shown by the green dashed line with error band and with the fit to the recent MINER$\nu$A antineutrino-hydrogen data~\cite{MINERvA:2023avz} shown by the blue dot-dashed line with error band. 
    }
     \label{fig:comparison}\vspace*{-0.6cm}
\end{figure}
The pion-nucleon form factor $G_{\pi NN}(Q^2)$ at the pion pole gives the pion-nucleon coupling 
$g_{\pi NN} \equiv \lim_{Q^2 \rightarrow -m_\pi^2} G_{\pi NN}(Q^2)= \lim_{Q^2 \rightarrow -m_\pi^2} (Q^2+m_\pi^2) G_P(Q^2)/4m_\pi^2F_\pi$, where $F_\pi$ is the pion decay constant.
Using the latter relation, we find $g_{\pi NN}= 13.25(67)(69)$.
 By extracting the nucleon matrix element of the pseudoscalar current we determine the pseudoscalar form factor $G_5(Q^2)$, which has a similarly $Q^2$ dependence as   $G_P(Q^2)$. 
  If PPD is satisfied then the ratio $r_{\rm PPD,2} \equiv \frac{4m_N}{m_\pi^2}\frac{m_qG_5(Q^2)}{ G_P(Q^2) }\bigg|_{Q^2\rightarrow -m^2_\pi} $ should be unity. This ratio is shown in Fig.~\ref{fig:ratios} and deviates from unity. From the slope one can compute the Goldberger-Treiman deviation $\Delta_{GT}$. We find $\Delta_{GT}=-0.0213(38)$ in agreement with the  2\% deviation predicted by chiral perturbation theory~\cite{Nagy:2004tp}  and determine the low energy constant $\bar{d}_{18}=g_A\Delta_{GT}/2m_\pi^2=-0.73(13)$ GeV$^{-2}$. Since PCAC is an exact operator relation, it provides a stringent test of our analysis on the form factor level. 
Therefore the ratio
 \begin{equation}
r_{\rm PCAC}(Q^2) = \frac{\frac{m_q}{m_N} G_5(Q^2) + \frac{Q^2}{4 m_N^2} G_P(Q^2) }{G_A(Q^2)} \,,
\end{equation}
should  be unity if lattice artifacts are correctly accounted for. In Fig.~\ref{fig:comparison}, we show $r_{\rm PCAC}$ for the  three ETMC ensembles of Table~\ref{tab:ens}. For finite lattice spacing the ratio is not unity. However, when we take the continuum limit the PCAC is satisfied as expected. In Fig.~\ref{fig:comparison} we compare our final value of $G_A(Q^2)$ in the continuum limit with that determined by the CLS-Mainz  collaboration. ETMC results are fully compatible with the results of CLS-Mainz and also favor the latest analysis of the MINER$\nu$A antineutrino-hydrogen data~\cite{MINERvA:2023avz}. \vspace*{-0.5cm}
 
\section{Decomposition of the nucleon spin} \vspace*{-0.2cm}
A surprising result was found  by the
European Muon Collaboration (EMC)~\cite{EuropeanMuon:1987isl,EuropeanMuon:1989yki} more than 30 years ago, namely that only about half the proton spin was carried  by its valence quarks. This triggered numerous experimental and theoretical studies  to understand  the so-called  proton spin puzzle. In  lattice QCD the total angular momentum  carried by valence and sea quarks and gluons can be computed by evaluating the second Mellin moments~\cite{Alexandrou:2021awv,Alexandrou:2020sml,Alexandrou:2017oeh}. We extract these moments by evaluating the nucleon matrix elements of the traceless part of the  energy momentum tensor $\bar{T}^{\mu \nu}_{q,g} $. They  can be decomposed into generalized form factors (GFFs) that depend only on the momentum transfer squared $q^2$. In Minkowski space we have~\cite{Ji:1998pc}
\begin{equation}
   \langle N(p',s') | \bar{T}^{\mu \nu}_{q,g} | N(p,s) \rangle = \bar{u}_N(p',s') \bigg[
    A_{20}^{q,g}(q^2) \gamma^{\{\mu} P^{\nu\}}
    + B_{20}^{q,g}(q^2) \frac{ i \sigma^{\{\mu \rho} q_\rho P^{\nu\}} }{2 m_N} 
    + C_{20}^{q,g}(q^2) \frac{q^{\{ \mu} q^{\nu\}}}{m_N}\bigg] u_N(p,s)
    \label{Eq:Decomp}
\end{equation}
where $u_N$ is the nucleon spinor with initial (final) momentum $p(p')$ and spin $s(s')$,  $P=(p'+p)/2$ is the total momentum and $q=p'-p$ the momentum transfer.  $A_{20}^{q,g}(q^2)$, $B_{20}^{q,g}(q^2)$ and $C_{20}^{q,g}(q^2)$ are the three GFFs.
In the forward limit, $A_{20}^{q,g}(0)$ gives the quark and gluon average momentum fraction $\langle x \rangle^{q,g}$. Summing over all quark and gluon contributions gives the momentum sum  $\langle x \rangle^{q} + \langle x \rangle^g = 1$. Furthermore,  the total spin carried by a quark is given by $J^q=\frac{1}{2}\left[A^q_{20}(0)+B^q_{20}(0)\right]$~\cite{Ji:1996ek}.

\begin{figure}[ht!]
  \begin{minipage}{0.48\linewidth}
  \includegraphics[width=\linewidth]{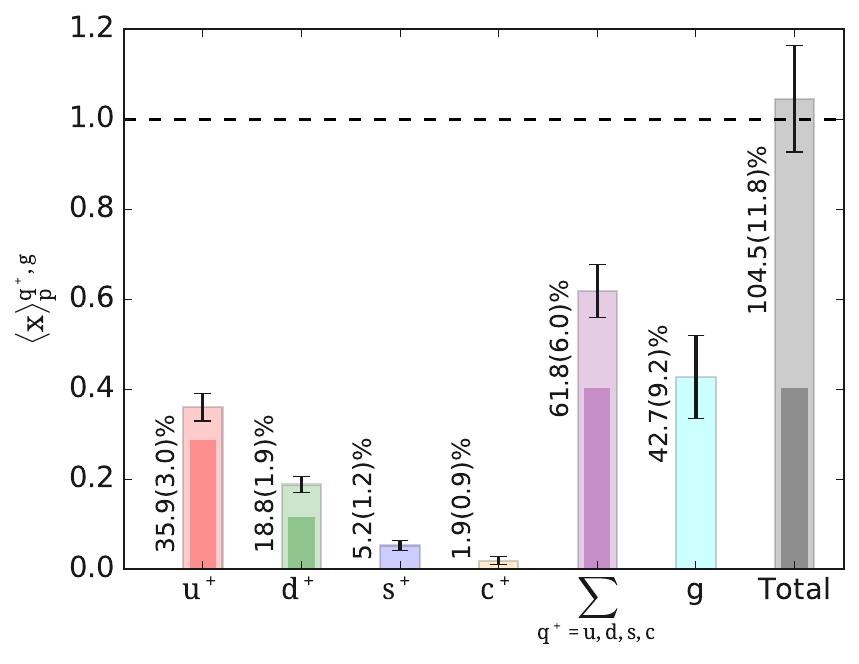}
  \end{minipage}\hfill
    \begin{minipage}{0.48\linewidth}
  \includegraphics[width=\linewidth]{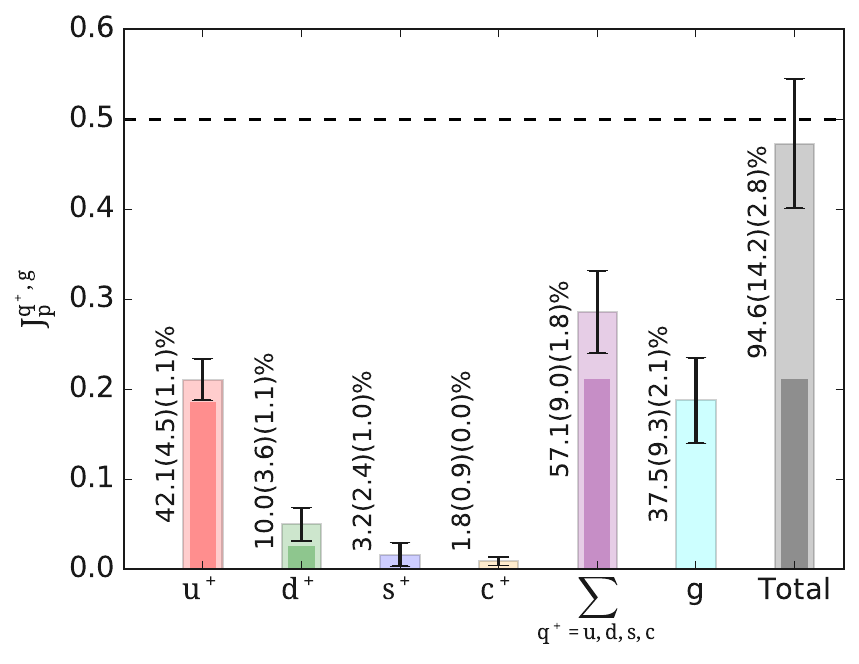}
\end{minipage}
 \caption{The decomposition of the proton average momentum fraction $\langle x \rangle_P$ (left) and  spin $J_P$. We show the contribution of  the up (red bar), down (green bar), strange (blue bar) and charm (orange bar) quarks and their sum (purple bar),  the gluon (cyan bar) and the total sum (grey bar). Note that what is shown  is the contribution of both the  quarks and antiquarks ($q^+=q+\bar{q}$). 
 Whenever two overlapping bars appear the darker bar denotes the purely connected contribution while the light one is the total contribution, which includes disconnected taking into account also the mixing. The error bars on the only connected part are omitted while 
 for the total are shown explicitly on the bars. The percentages written in the figure are for the total contribution.
 The dashed horizontal line is the momentum and spin sums. Results are given in $\mathrm{ \overline{MS}}$ scheme at 2~GeV.}\vspace*{-0.7cm}
  \label{fig:averXBar}
 \end{figure}

In Fig.~\ref{fig:averXBar}, we show  our results for the proton average momentum
fraction for the up, down, strange and charm quarks, for the gluons as well as their sum.
The up quark makes the largest quark contribution of about 35\% and it is twice as big as that of the down
quark. The strange quark contributes significantly smaller, namely about 5\% and the charm contributes about 2\%.
The gluon has a significant
contribution of about 45\%.  Summing all the contributions 
results to
$
\sum_{q=u,d,s,c}\langle x \rangle_R^{q^+} + \langle x\rangle_R^g = 1.045(118)$
confirming the expected momentum sum.
Fig.~\ref{fig:averXBar}, showing connected and disconnected contributions,  demonstrates that disconnected contributions are crucial and if  excluded 
would result to a significant underestimation of the momentum sum.  

The individual contributions to the proton spin are presented in Fig.~\ref{fig:averXBar}. The major contribution comes from the up quark amounting to about 40\% of the proton spin. The down, strange and charm quarks have relatively smaller contributions. All quark flavors together constitute to about 60\% of the proton spin. The gluon contribution is as significant as that of the up quark,  providing the missing piece to obtain $J_P=94.6(14.2)(2.8)$\%  of the proton spin, confirming indeed the spin sum.
The $\sum_{q=u,d,s} B_{20}^{q^+}(0) + B_{20}^g(0)$ is expected to vanish in order to respect the momentum and spin sums. We find for the renormalized values that $
    \sum_{q=u,d,s} B_{20,R}^{q^+}(0) + B_{20,R}^g(0)=-0.099(91)(28)$,
which is indeed compatible with zero. We note that these results were obtained using the ETMC ensemble CB211.072.64 and are thus not extrapolated to the continuum limit. The computation using the other two ensembles of Table~\ref{tab:ens} is ongoing.  \vspace*{-0.5cm}
\section{Direct computation of PDFs and GPDs} \vspace*{-0.2cm}
 PDFs are light-cone correlation matrix elements given by
 \begin{equation}
   F_\Gamma(x)=\frac{1}{2}\int\frac{dz^-}{2\pi}\,e^{ixP^{+}z^{-}}\langle N(p)|\bar{\psi}(-z/2)\Gamma W(-z/2,z/2)\psi(z/2)|N(p)\rangle|_{z^+=0,\vec{z}=0},
   \label{PDFs}
 \end{equation}
 and as such cannot be computed on a Euclidean lattice.
\begin{figure}[h!]
    \centering
    \includegraphics[width=0.46\linewidth]{isov_Helicity.pdf}
    \includegraphics[width=0.48\linewidth]{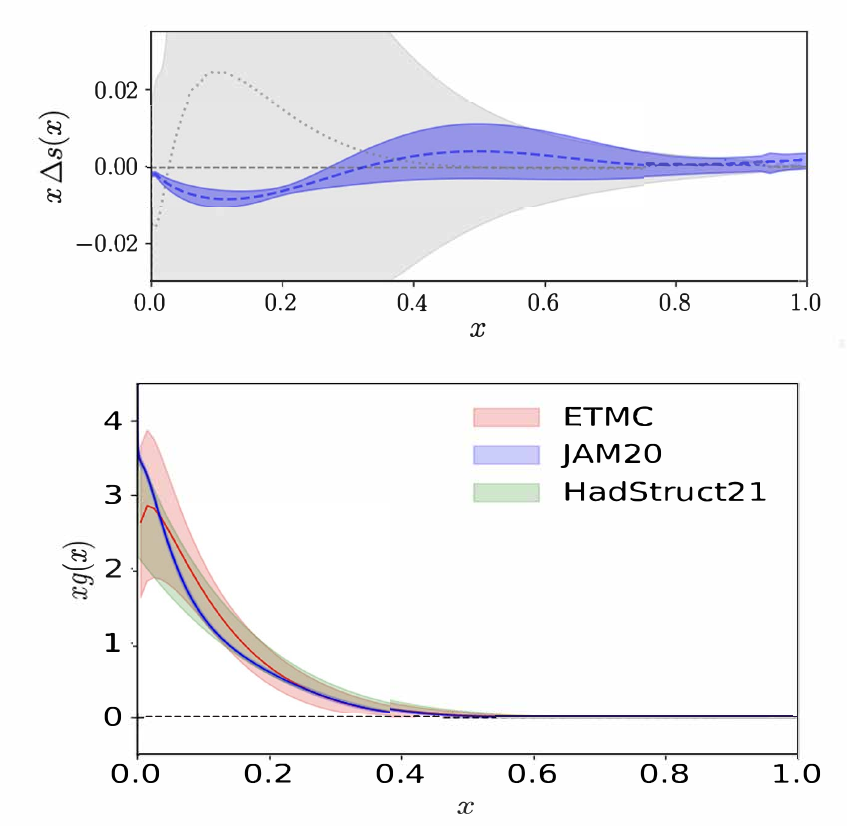} \vspace*{-0.3cm}
    \caption{Left:  JAM17 results are shown
with the red band, lattice QCD results are shown with the yellow band~\cite{Alexandrou:2018pbm} using one ETMC ensemble with physical pion mass, while the blue band shows the combined fits using both lattice and experimental data~\cite{Bringewatt:2020ixn}. Right top: Strange helicity using an ETMC ensemble with  pion mass of 260~MeV (blue band) compared to phenomenological analysis (gray band)~\cite{Alexandrou:2020uyt}. Right bottom:  A comparison of ETMC results on the gluon PDF~\cite{Delmar:2023agv} (red) with those of of HadStruc~\cite{HadStruc:2021wmh} (green), and the global analysis of JAM20~\cite{Moffat:2021dji} (blue). Results are shown in the MS scheme at a scale of 2 GeV. }
    \label{fig:helicity} \vspace*{-0.3cm}
\end{figure}
 In a pioneering paper,  X. Ji\cite{Ji:2013dva} proposed instead to compute matrix elements of spatial correlators e.g. along the $z$-axis, and a  nucleon state with a large momentum boost in the $z$-direction. For large enough boosts, one can relate these so called quasi-distributions to the light-cone distributions through a matching kernel computed in perturbation theory. Within this large momentum effective theory (LaMET) one can then extract directly PDFs and, allowing momentum transfers, the GPDs. For PDFs, we compute
 \begin{equation}
  \tilde{F}_\Gamma(x,P_3,\mu) =2P_3
  \int_{-\infty}^{\infty}\frac{dz}{4\pi}e^{-ixP_3z}\,\langle P_3\vert\,\overline{\psi}(0)\, \Gamma W(0,z)\, \psi(z)|\,P_3\rangle |_{\mu}
 \end{equation}
 and, after non-perturbative renormalization, we match the quasi-PDF $\tilde{F}_\Gamma(x,P_3,\mu)$ to extract the PDF
   $$\tilde{F}_\Gamma(x,P_3,\mu)=\int_{-1}^1
   \frac{dy}{|y|} \, C\left(\frac{x}{y},\frac{\mu}{yP_3}\right)\, {F}_\Gamma(y,\mu) +{\cal{O}}\left(\frac{m_N^2}{P_3^2},\frac{\Lambda^2_{\rm QCD}}{P_3^2}\right)$$
   through the matching kernel $C\left(\frac{x}{y},\frac{\mu}{yP_3}\right)$.
   Although these computations are not as mature as the computation of Mellin moments, first results are very promising. In Fig.~\ref{fig:helicity}, we show results on the nucleon isovector helicity from lattice QCD,  from the JAM collaboration analysis, and when  combined. As can be seen, using lattice QCD input greatly improves the theoretical predictions. 
In Fig.~\ref{fig:helicity}, we also show the gluon PDF computed using pseudo-distributions, a variant of quasi-distributions. \vspace*{-0.5cm}

\section{Conclusions} \vspace*{-0.2cm}
   
 Lattice QCD results produce known experimental values of  e.g. the nucleon axial charge, and the electromagnetic form factors and, thus, can be reliably used to predict other less known quantities such as the tensor charge, axial form factors, and pseudoscalar form factor providing valuable input to experiments and phenomenology. 
Second Mellin moments that probe the distribution of spin among the quarks and gluons can be extracted reliably and provide a quantitative understanding of the fraction of the momentum and spin carried by quarks and gluons on the proton. 
Direct computation of PDFs, GPDs and TMDs can now be computed in lattice QCD and a lot of progress is foreseen in the near future on determining these important quantities  that will   provide a more complete picture of hadron structure.

\medskip
\noindent
{\bf Acknowledgments.} C.A. acknowledges partial support by the project 3D-nucleon, id number EXCELLENCE/0421/0043, co-financed by the European Regional Development Fund and the Republic of Cyprus through the Research and Innovation Foundation and by the European Joint Doctorate AQTIVATE that received funding from the European Union’s research and innovation program under the Marie Skłodowska-Curie Doctoral Networks action and Grant Agreement No 101072344.   S.B. is funded by the project QC4LGT, id number EXCELLENCE/0421/0019, co-financed by the European Regional Development Fund and the Republic of Cyprus through the Research and Innovation Foundation. \vspace*{-0.5cm}
% BibTeX or Biber users please use (the style is already called in the class, ensure that the "woc.bst" style is in your local directory)
% \bibliography{name or your bibliography database}

\end{document}